\documentclass[prl,twocolumn,showpacs,preprintnumbers,amsmath,amssymb,superscriptaddress]{revtex4}
\usepackage{graphicx}% Include figure files
\usepackage{graphics}% Include figure files
\usepackage{dcolumn}% Align table columns on decimal point
\usepackage{bm}% bold math

\begin{document}

\title{Magnetic excitations in NpCoGa$_5$}% Force line breaks with \\

\author{N. Magnani}
\affiliation{European Commission, Joint Research Centre, Institute for
Transuranium Elements, Postfach 2340, D-76125 Karlsruhe, Germany}

\author{A. Hiess}
\affiliation{Institut Laue Langevin, BP 156, F-38042 Grenoble Cedex 9, France}

\author{R. Caciuffo}
\affiliation{European Commission, Joint Research Centre, Institute for
Transuranium Elements, Postfach 2340, D-76125 Karlsruhe, Germany}

\author{E. Colineau}
\affiliation{European Commission, Joint Research Centre, Institute for
Transuranium Elements, Postfach 2340, D-76125 Karlsruhe, Germany}

\author{F. Wastin}
\affiliation{European Commission, Joint Research Centre, Institute for
Transuranium Elements, Postfach 2340, D-76125 Karlsruhe, Germany}

\author{J. Rebizant}
\affiliation{European Commission, Joint Research Centre, Institute for
Transuranium Elements, Postfach 2340, D-76125 Karlsruhe, Germany}

\author{G. H. Lander}
\affiliation{European Commission, Joint Research Centre, Institute for
Transuranium Elements, Postfach 2340, D-76125 Karlsruhe, Germany}

\date{\today}% It is always \today, today,

\begin{abstract}
We report the results of inelastic neutron scattering experiments on 
NpCoGa$_{5}$, an isostructural analogue of the PuCoGa$_{5}$ 
superconductor. Two energy scales characterize the magnetic response 
in the antiferromagnetic phase. One is related to a non-dispersive 
excitation between two crystal field levels. The other at lower 
energies corresponds to dispersive fluctuations emanating from the 
magnetic zone center. The fluctuations persist in the paramagnetic 
phase also, 
although weaker in intensity. This supports the possibility that magnetic
fluctuations are present in PuCoGa$_{5}$, where unconventional $d$-wave 
superconductivity is achieved in the absence of magnetic order.
\end{abstract}

% PACS, the Physics and Astronomy Classification Scheme.
%\keywords{Suggested keywords}%Use showkeys class option if keyword display desired
\maketitle

Many complex and diverse phenomena have been observed in the series 
of isostructural RTX$_{5}$ intermetallic compounds containing cerium 
or a light actinide element R, a transition metal T (Fe, Co, Ni, Rh, 
Ir), and an element of the boron group X (Ga or In). Current research 
on this family is enhanced by the possibility to explore links between
unconventional superconductivity, magnetism, and non-Fermi liquid
behavior in proximity of a quantum critical point \cite{sarrao07}.
Especially the discovery of $d$-wave superconductivity in 
PuCoGa$_{5}$ at the surprisingly high critical temperature of 
T$_{sc}$ = 18 K \cite{sarrao02} raised much interest. No long-range 
magnetic order is observed in PuCoGa$_{5}$, but theoretical
work \cite{opahle03,maehira03,shick05,pourovskii06} as well as 
recent NMR experiments \cite{curro05} suggest the presence of spin 
polarization and hence strong magnetic correlations. Evidence for non 
$s$-wave superconductivity is reported also for CeRhIn$_{5}$ and 
CeCoIn$_{5}$, where heavy fermion character is consistent with 
proximity to magnetic order \cite{sarrao07}.

Although magnetic fluctuations have been proposed to stabilize 
superconductivity for both Pu and Ce-based compounds, no direct 
measurement of magnetic fluctuation spectra has yet been reported. 
This is mainly because of the large neutron absorption cross-section 
of $^{239}$Pu and $^{113}$In, and the lack of large single crystals 
of $^{242}$PuCoGa$_{5}$. Thus, at present, there is no information on 
the electronic excitations in this interesting set of materials.

Here we present the results of inelastic neutron scattering (INS)
experiments on the neptunium-based analogue NpCoGa$_{5}$. This 
compound (space group P4/mmm, $a$ = 4.277~\AA, $c$ = 6.787~\AA) 
shows antiferromagnetic (AF) order below T$_{N}$ = 47 K, with an ordered 
Np moment $\mu = 0.84 \mu_{B}$ \cite{colineau04} pointing along the 
tetragonal c-axis \cite{metoki05}. The magnetic structure is defined 
by the propagation vector ($0~0~1/2$). This is exactly the reciprocal 
space position where theory predicts the dominant weight for the 
magnetic fluctuations in PuCoGa$_{5}$ \cite{pourovskii06}. We 
therefore expect some similarities between the latter and the magnetic
excitations in the antiferromagnetic compound NpCoGa$_{5}$.

For this experiment we used a large single crystal with a mass of 
1.1~g, grown and characterized at the Institute for Transuranium 
Elements, Karlsruhe. Measurements were performed with the thermal 
triple axis spectrometer IN8 at the Institut Laue-Langevin, Grenoble,
using a Si($1~1~1$) focussing monochromator, open collimation, two 
graphite filter for second-order contamination suppression, and a 
pyrolytic graphite ($0~0~2$) analyzer.

The final wave-vector was fixed at $k_{f}$ = 2.662 \AA$^{-1}$. 
The crystal was oriented with the $a$- and $c$-axis in the scattering 
plane. Several structural Bragg peaks were measured to test its 
quality, found to be adequate for an INS experiment. The rocking 
curve ($\Theta$ sample rotation) at the ($0~0~3$) Bragg reflection shows a 
double-peak structure, the two maxima being separated by about 1.6 
degrees and each peak having 0.9 degrees full width at half maximum 
(FWHM). The sample orientation was optimized for the crystallite 
with largest volume. Longitudinal scans at the ($2~0~0$) and 
($0~0~3$) Bragg peaks gave $\Delta q_{h}$ = 0.026 and $\Delta q_{l}$ = 
0.045~r.l.u. FWHM, respectively.

\begin{figure}
\includegraphics[width=7.5cm]{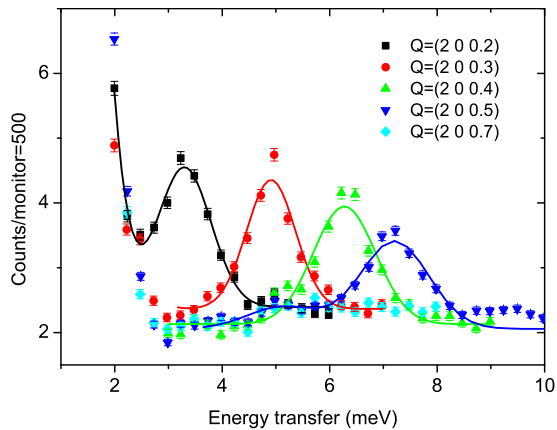}
\caption{Neutron groups observed with the sample at 10 K in
constant-Q measurements at different wave
vectors ($2~0~q_z$). The
observed excitations correspond to TA phonons propagating along
the [$0~0~1$] direction. \label{phonons}}
\end{figure}

Before searching for the magnetic inelastic response, we measured 
the neutron groups corresponding to the transverse acoustic (TA) 
phonons propagating along [$0~0~1$] to determine the lowest 
possible vibrational excitation energy at the magnetic zone center 
($h~k~l+1/2$). Constant-Q scans were performed around the ($2~0~1/2$) 
reciprocal lattice position with the sample kept at T = 10 K. The 
results obtained are shown in Fig. \ref{phonons}. The observed energies and
intensities are in good agreement with the results reported for
UCoGa$_{5}$ \cite{metoki06}. In particular, the intensity at ($2~0~0.7$)
is very weak in all cases. Since the TA[$0~0~1$] phonon branch reaches
an energy of 7.2~meV at the magnetic zone center, any excitation with 
energy smaller than this value cannot be of phononic origin. 

\begin{figure}
\includegraphics[width=7.5cm]{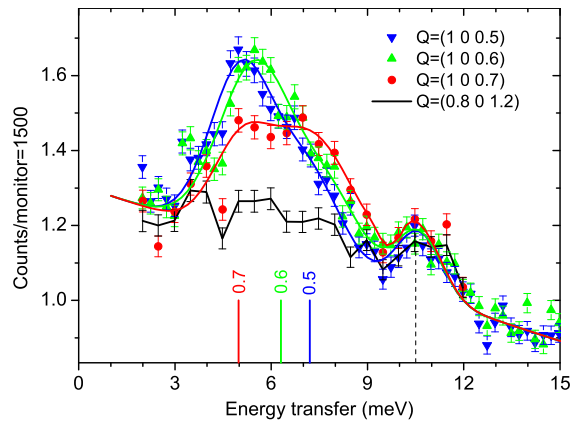}
\caption{Neutron groups corresponding to excitations propagating
along [$0~0~1$] at 10 K.
The solid lines are fits to the sum of two Gaussians and a
sloping background, measured at
Q = ($0.8~0~1.2$). The third peak at 10.5~meV
(dashed vertical line) is
attributed to a local, non dispersive excitation. The vertical tags
indicate the energy of the TA phonons, and are labelled by 
the corresponding $q_{z}$ value.
\label{magdisp}}
\end{figure}

Having established the energy of the lowest vibrational excitations, 
we looked for the magnetic response at the ($1~0~1/2$) reciprocal 
lattice point. At this position, the intensity of the TA phonon is 
expected to be weaker than at ($2~0~1/2$) by almost an order of magnitude,
because of the geometric factor appearing in the neutron cross-section.
The results obtained are shown in Fig. \ref{magdisp}. Data were taken as
constant-Q scans at different Q = ($1~0~q_z$) positions, and fitted to
two Gaussian line-shapes. One Gaussian curve represents the 
contribution of the TA phonon; it is centered at the energy shown in
Fig. \ref{phonons} for the corresponding peak. The intensity
scales with the square of the neutron momentum transfer, as 
expected. The second Gaussian curve is attributed to a
propagating magnetic excitation, with an energy gap of $5.5 \pm
0.2$ meV. Moving away from the magnetic zone center ($1~0~1/2$), the
energy of phonons decreases whereas the energy of the magnetic peak 
increases. 

In addition, a weaker non-dispersive signal is observed at about
10.5~meV, which we attribute to a mean field excitation between 
crystal field (CF) levels. Due to its non dispersive behavior, this
excitation is also visible in the scan performed at a generic
reciprocal lattice position ($0.8~0~1.2$), as shown in Fig. 
\ref{magdisp}. 

Equivalent scans at ($2~0~1/2$) and ($3~0~1/2$) establish an increase 
in the phonon contribution only, and confirm the magnetic origin of 
both,  the propagating and the local excitation.

\begin{figure}
\includegraphics[width=7.5cm]{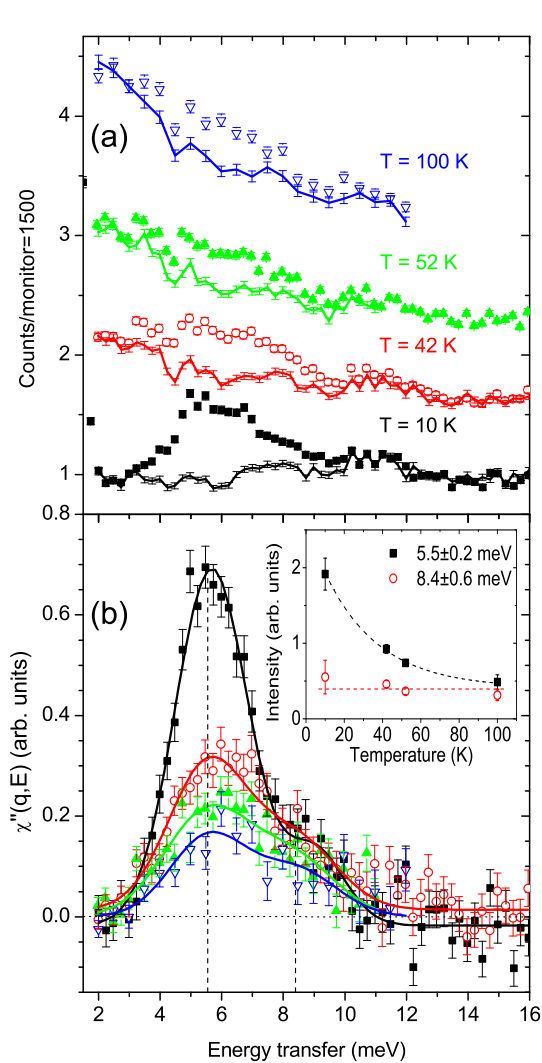}
\caption{(a) Constant-Q scans at different temperatures at the Q =
($0~0~3/2$). A constant vertical offset of 1000 counts is
applied for clarity. The solid lines are scans at Q = 
($0.2~0~1.47$). (b) Imaginary component of the dynamical magnetic susceptibility: squares 10
K, open circles 42 K, up triangles 52 K, down open triangles 100 K.
The solid lines are fit to two Gaussian line-shapes centered at 5.5~$\pm$~0.2
and 8.4~$\pm$~0.6~meV (vertical dashed lines), with intensities shown in the inset.
\label{magtemp}}
\end{figure}

To better characterize the non-local magnetic response, we followed 
its temperature dependence by performing constant-Q scans at the ($0~0~3/2$) 
reciprocal lattice point (Fig. \ref{magtemp}). At this position, the TA phonon 
contribution is minimized and the longitudinal acoustic (LA) phonon is 
expected at higher energies only. In addition, at such a small 
momentum transfer, all phononic signals are weak. The scan at the 
AFM zone center is compared with a scan at Q = ($0.2~0~1.47$), which
was reached by a $\Theta$ = 12.5 degrees sample rotation away from 
the optimal value for the ($0~0~3/2$) position. This position might 
be considered to reflect the Q-independent background (see discussion 
below). By subtracting the background from the intensity at the AFM zone 
center and dividing by the Bose factor, we obtain the temperature 
dependence of the imaginary component of the dynamical 
susceptibility (Fig. \ref{magtemp} b). A phonon contribution is 
visible at 8.4~$\pm$0.6~meV and, as expected after Bose factor
normalization, its intensity is temperature independent. On the other 
hand the magnetic intensity at 5.5~meV becomes weaker as the 
temperature increases (inset Fig.\ref{magtemp}), but it is still 
clearly visible above the N\'{e}el temperature, T$_{N}$. The large 
difference between the intensities measured at
($0~0~1.5$) and ($0.2~0~1.47$) confirms that this signal
cannot be attributed to a crystal field excitation, neither below nor 
above the N\'{e}el temperature T$_{N}$. Indeed, the intensity of a 
crystal field excitation would change only as the square of the 
magnetic form factor, which is essentially the same 
at the two positions.

It is interesting to compare the static and dynamic magnetic 
intensities at both the ($1~0~1/2$) and the ($0~0~3/2$) positions. 
The static magnetic moment points along the tetragonal c-axis, and 
therefore all Bragg reflections of ($0~0~L$)-type have zero neutron 
intensity. Since the inelastic magnetic intensity  at 5.5 meV of 
both reflections is comparable, we suspect the dynamic 
fluctuations to show an (at least) significant transverse component.

\begin{figure}
\includegraphics[width=7.5cm]{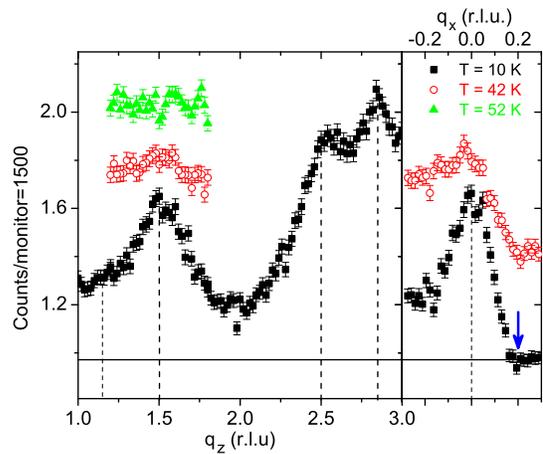}
\caption{Q scans at constant energy transfer, $\hbar \omega$ = 5.5~meV.
Left panel: longitudinal scan along the [$0 0 1$]
direction, Q = ($0~0~q_{z}$). Right panel: transverse scan
around ($q_{x}~0~1.5$), performed by $\Theta$ rotation of the sample.
Neutron intensity is plotted as a function of $q_{x}$.
The arrow indicates the position corresponding to the scan at Q = 
($0.2~0~1.47$) in Fig. \ref{magtemp}. A vertical offset of 
0.2 counts/monitor is applied for clarity to data at different temperature. 
\label{Qscan}}
\end{figure}

Fig.\ref{Qscan} shows longitudinal Q-scans around Q = ($0~0~q_{z}$) and 
transverse Q-scans around Q = ($q_{x}~0~3/2$), with neutron energy-transfer
fixed at 5.5~meV. The non-local character of the 5.5~meV excitation 
at T = 10~K is confirmed by the strong Q dependence of the
intensity, which peaks at the magnetic zone center. The peaks at
($0~0~2.85$) and (very much weaker) at ($0~0~1.15$) are due to the 
nearby LA phonon.  At this stage of our investigations, with a relaxed 
instrumental momentum resolution and without polarized neutrons, we 
cannot establish the origin of the large background signal, which 
increases  with temperature. This prevents any quantitative 
analysis of the peak width and therefore the spatial coherence length 
of the dispersive magnetic excitation. We note, however, that at 10~K 
the magnetic peak has a width comparable to that of the LA phonon,
suggesting a long-range character. Approaching T$_{N}$, the magnetic
response becomes much broader in Q, and only a short-range in-plane correlation
appear to survive. Polarization analysis experiments are planned to address
this point.

\begin{figure}
\includegraphics[width=7.5cm]{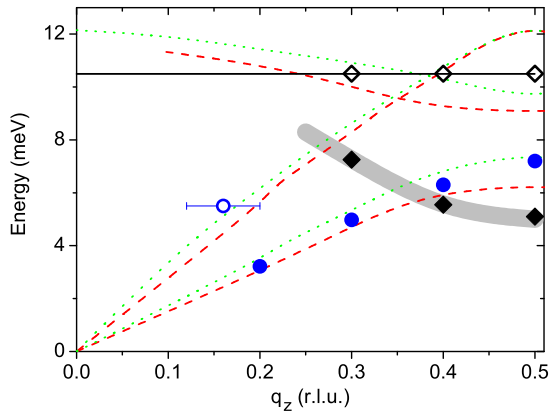}
\caption{Dispersion relation along [$0~0~1$] for the TA phonon (full
circles), the LA phonon (open circle), and the two magnetic
modes
(full and open diamonds). Data taken at T = 10~K. Dashed and dotted
lines are phonon dispersion curves reported for PuCoGa$_{5}$ and UCoGa$_{5}$,
respectively \cite{raymond06,metoki06}.
\label{disp}}
\end{figure}

Fig. \ref{disp} shows the dispersion relation along [$0~0~1$] of 
the excitation modes observed at 10~K in this experiment. The TA phonon dispersion 
falls in between the dispersion curves reported for the
Pu and U homologous compounds \cite{raymond06,metoki06}. Also the 
point measured for the LA phonon is in good agreement with previous 
studies of the vibrational response in UCoGa$_{5}$ and PuCoGa$_{5}$.
The dispersion of the low-energy magnetic mode is
typical of a spin-wave excitation in an antiferromagnet. The 
non-dispersive local excitation at 10.5 meV is also shown. 

Previous studies \cite{aoki04} assume the 5f electrons to be 
essentially localized and predict a crystal field (CF) level at about 
8 meV in the paramagnetic phase, directly coupled with a doublet
ground state. On this basis, an Ising mean field (MF) model has been
proposed to describe the magnetically ordered phase \cite{kiss06}. The
Hamiltonian can be written as the sum of a CF term,

\begin{equation}
H_{CF} = B^{0}_{2} O^{0}_{2} + B^{0}_{4} O^{0}_{4} + B^{0}_{6}O^{0}_{6} + B^{4}_{4}O^{4}_{4}
+ B^{4}_{6}O^{4}_{6}
\end{equation}

a nearest-neighbors exchange interaction within the $a$-$b$ planes,

\begin{equation}
H_{ab} = - \frac{1}{4} \Lambda_{z} \sum_{<ij>}J_{z, i}^{(l)}J_{z,j}^{(l)} 
\end{equation}

and an interlayer exchange term

\begin{equation}
H_{c} = \frac{1}{2} \Lambda_{z}^{\perp} \sum_{l}J_{z, i}^{(l)}J_{z,
i}^{(l+1)},
\end{equation}

where $O_{k}^{q}$ are Stevens operator equivalents, $B_{k}^{q}$ the CF
parameters, $\Lambda_{z}$ and $\Lambda_{z}^{\perp}$ are the intralayer
and interlayer exchange
constants, respectively, $i, j$ label the Np site in the $a$-$b$ plane, 
and $l$ indexes the layers. 

Following Aoki et al. \cite{aoki04}, the lowest lying CF eigenstates have the form

\begin{eqnarray}
|d_{\pm}\rangle &=& a|\pm 3\rangle + \sqrt{1-a^{2}}|\mp 1\rangle
\nonumber \\
|s\rangle &=& c(|+ 4\rangle + |- 4\rangle) + \sqrt{1-2c^{2}}|0\rangle.
\end{eqnarray}

The doublet $|d_{\pm}\rangle$ and the singlet $|s\rangle$ are separated by a
gap $\Delta_{CF}$, whilst the remaining states of the $^{5}I_{4}$
multiplet have much higher energies and do not affect the low
temperature properties. The exchange Hamiltonian is treated in MF
approximation and the self-consistent solution is sought.
The coefficient $a$ is fixed by the value of
the ordered moment, and the sum $\Lambda_{z} + \Lambda_{z}^{\perp}$ is 
varied in order to fix T$_{N}$ to the experimental value. A 
non-dispersive INS transition at 10.5~meV is indeed allowed if $\Delta_{CF}$
= 6~meV. This bare CF gap is close to the value
proposed in \cite{aoki04} to explain the bulk properties in the
paramagnetic phase. In the ordered phase, the doublet would be split by
$\Delta_{ex}$ = 9~meV, however the corresponding INS transition would not 
be dipole-allowed. Therefore, within this model, the dispersive excitation appearing at
5.5~meV at the magnetic zone center cannot be understood. Its observation 
indicates that a fully localized dipolar model fails to describe the dynamical 
magnetic response of NpCoGa$_{5}$. 

In conclusion, our inelastic neutron scattering experiment gives
evidence for the existence of two energy scales for the magnetic 
excitations in the ordered phase of
NpCoGa$_{5}$. One is related to a non-dispersive excitation between two crystal
field levels (with a bare gap modified by mean-field exchange), and a 
lower one corresponds to dispersive spin fluctuations.
The CF splitting is compatible with earlier predictions based on
essentially localized dipolar models, which reproduce the bulk
magnetic properties in the paramagnetic phase \cite{aoki04}. The dispersive
magnetic response, on the other hand, reveals the presence of non-local
excitations with an energy gap very close to that deduced from the low
temperature dependence of the specific heat \cite{colineau04}.

As shown in Fig. \ref{magtemp}, an inelastic magnetic response
is visible also at 52 and 100~K. This signal cannot be attributed to a
CF transition, because of the different intensities measured at Q =
($0~0~3/2$) and at Q = ($0.2~0~1.47$). Therefore, although broader 
and weaker, the low energy excitation resulting from the spin-spin
temporal correlation clearly persist in the paramagnetic phase. This 
supports the possibility that magnetic
fluctuations centered around ($0~0~1/2$) are present, as suggested by
theory \cite{pourovskii06}, also in
PuCoGa$_{5}$, where they have been proposed as a possible
mediating boson responsible for the appearance of superconductivity \cite{bang04}.

Finally, it appears that the $5f$ electrons in NpCoGa$_{5}$ are neither 
completely localized nor itinerant. This dual nature is typical of the
$5f$ shell in intermetallic compounds and its elucidation is one of
the main current topics in the physics of strongly correlated electron
systems. Recent M\"{o}ssbauer spectroscopy studies of NpTGa$_{5}$ point to a
tendency towards localization of the $5f$ states with increasing
number of $d$ electrons on the transition metal T \cite{colineau07}.
It will therefore be interesting to extend our INS investigation to
NpFeGa$_{5}$, where an essentially itinerant response is expected, and
to NpNiGa$_{5}$, where the localized dynamics should dominate.

\bibliography{NpCoGa5}

\end{document}